\documentclass[a4paper,11pt,reprint,showkeys]{revtex4}
\usepackage{url}
\usepackage{ulem}
\usepackage{amsmath, amssymb}
\usepackage{graphicx}

\begin{document}
\title{BABYSCAN -- a whole body counter for small children in Fukushima}
\author{Ryugo S. Hayano}
\thanks{Correspondence should be addressed: R. Hayano, (\url{hayano@phys.s.u-tokyo.ac.jp}).}
\affiliation{Department of Physics, The University of Tokyo, 7-3-1 Hongo, Bunkyo-ku, Tokyo 113-0033, Japan}
\author{Shunji Yamanaka}
\affiliation{Department of Mechanical and Biofunctional Systems, Institute of Industrial Science, The University of Tokyo, 4-6-1 Komaba, Meguro-ku, Tokyo 153-8505, Japan }
\author{Frazier L. Bronson and Babatunde Oginni}
\affiliation{Canberra Industries, Inc., 
800 Research Par∂kway, Meriden, CT 06450, U.S.A.}
\author{Isamu Muramatsu}
\affiliation{Canberra Japan KK,
4-19-8 Asakusabashi, Taito-ku, Tokyo 111-0053, Japan}

%\setstcolor{red}

\begin{abstract}
BABYSCAN, a whole body counter for small children with a detection limit for $^{137}$Cs of better than 50 Bq/body, was developed, and the first unit has been installed at a hospital in Fukushima, to help families with small children who are very much concerned about internal exposures. The design principles, implementation details and the initial operating experience are described.
\end{abstract}
\maketitle

\section{Introduction}

The Fukushima Dai-ichi NPP accident~\cite{tanaka} contaminated the soil of densely-populated regions of Fukushima Prefecture with radioactive cesium, which poses risks of internal (and external) exposures to the residents. If we apply the knowledge of post-Chernobyl accident studies~\cite{unscear1988}, internal exposures in excess of several mSv y$^{-1}$ would be expected to be frequent in Fukushima. 

Extensive whole-body-counter surveys of 21,785 residents in highly-affected Fukushima municipalities, however, showed that their actual internal exposure levels are much lower than estimated~\cite{PJAB}; in 2012--2013, the $^{137}$Cs detection percentages (the detection limit being $\sim 300$ Bq/body) are about 1\% for adults, and practically 0\% for children (age 6--15). These results are consistent with those of many other measurements and studies conducted so far in Fukushima, e.g., dose assessment~\cite{nagataki,tsubo1,tsubo2,sugimoto}, rice inspection~\cite{rice}, foodstuff screening and duplicate-portion studies~\cite{coop}.

Nevertheless, there continue to be many residents, families with small children in particular, who are very much concerned about internal exposures. This is in part due to the fact the whole body counters currently being used in Fukushima, such as the FASTSCAN~\cite{FASTSCAN}, are designed for radiation workers, who are adults.  Children have been successfully measured previously at Chernobyl, and in Fukushima Prefecture, by having them stand on a small stool to get their bodies into the detection zone.  While this is suitable to measure larger uptakes in larger children, it is not optimum for measuring small children ($\lesssim 4$ y), and is not suitable for infants or children who cannot stand.

Scientifically, it is sufficient to measure parents, but worried parents strongly request to have their babies measured.
We therefore launched a project in the spring of 2013~\cite{FB} to develop a whole body counter for small children called a ``BABYSCAN'', and have installed the first unit at the Hirata Central Hospital in Fukushima Prefecture in December 2013. The design principles, implementation and the initial operating experience are reported.

\section{BABYSCAN Requirements}

About 80\% of some 60 whole body counters currently installed in Fukushima Prefecture are Canberra's FASTSCAN. A subject stands for two minutes in a shielding box made of iron, which houses two $7.6 \times 12.7 \times 40.6$ cm sodium iodide (NaI) gamma-ray detectors.  The detection limit for radioactive cesium is about 250-300 Bq/body (both for $^{134}$Cs and $^{137}$Cs), which is nearly independent of height and/or weight of the adult subject (flat within $\sim \pm 15\%$).

This detection limit is however too high for reliably measuring small children, since the biological half-life of radioactive cesium in children ($\sim 13$ days for 1-year old, $\sim 30$ days for 5-year old) is much shorter than that in adults ($\sim 110$ days)~\cite{icrp78,mondal3}. As a result, children's internal contamination is harder to detect.

For example, 
if an adult ingested 3 Bq of $^{137}$Cs every day, the body burden would reach an equilibrium plateau of $\sim 400$~Bq/body~\cite{mondal3}. This can be detected by the FASTSCAN.
If on the other hand a 1-year-old child ingested the same amount, the resultant  body burden would be $\sim 60 $Bq/body. Therefore, the whole body counters for babies must have a much lower detection limit. 

Our goal was to achieve a detection limit of $< 50$ Bq/body for $^{134, 137}$Cs.
In order to realize this high sensitivity, the BABYSCAN must be ergonomically designed so that a small child can stay still for several minutes, without feeling afraid of confinement.

From the beginning, it was recognized that the BABYSCAN's design must be reassuring to parents, and that in addition to being a measurement device, it would be expected to play an important role as a communication tool to facilitate interactions between medical staff and residents.

\begin{figure*}[htbp]
\includegraphics[width=0.48\textwidth]{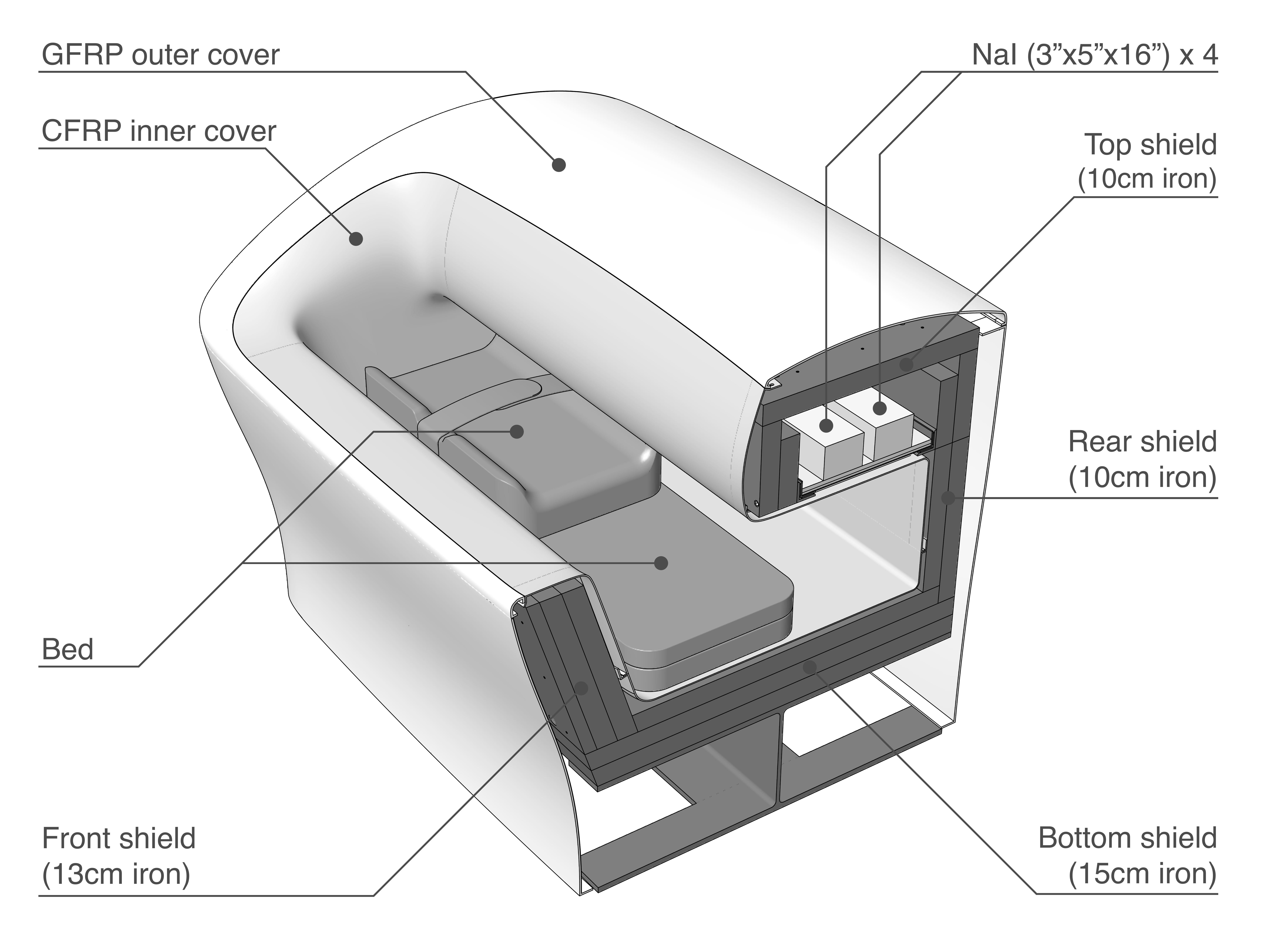}
\includegraphics[width=0.48\textwidth]{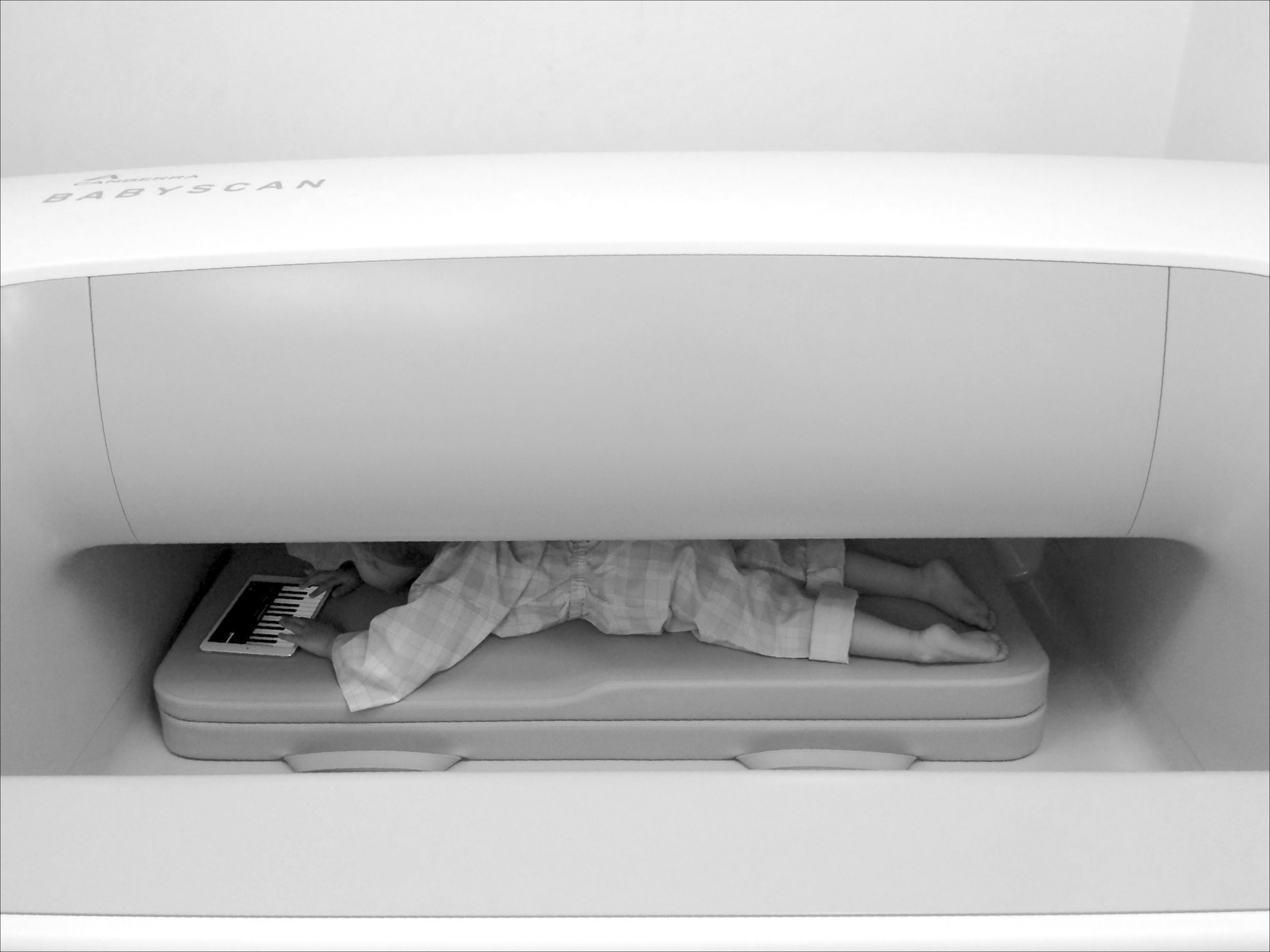}
\caption{\label{fig:cutaway} Left: a cutaway view of the BABYSCAN. Right: a 4-year-old child lying on front, playing with a tablet computer, during a 4-minute measurement in the BABYSCAN.}
\end{figure*}

\section{BABYSCAN Details}

The BABYSCAN's design principles and technologies were derived from those of FASTSCAN, but in order to realize higher sensitivity, there are some crucial differences.

As shown in Fig.~\ref{fig:cutaway}, the subject lies down inside the measurement chamber of BABYSCAN, as opposed to standing as in the case of FASTSCAN. A child can either lie on the bed supine (on their back and face up), or prone (on their stomach, face down). During development, we discovered that older children's posture tends to be more stable in the prone position, as shown in the right-hand panel of Fig.~\ref{fig:cutaway}. Small babies, however, tend to prefer the supine position, and are more comfortable when they can also see their mother's face through the opening. Both positions are OK, as they have essentially the same efficiency.

There are four NaI detectors ($7.6 \times 12.7 \times 40.6$ cm each), arranged in a two-by-two geometry, installed in an iron-shielded compartment placed above the subject. The bottom of the NaI compartment has a window facing the subject, made of a carbon-honeycomb plate measuring 28 cm $\times$ 86 cm. 
The detectors are close to each other, therefore the area of the NaI group of detectors is approximately 26 cm $\times$ 82 cm. This is similar to the size of a small child, thereby achieving a high gamma-ray detection efficiency.

\begin{table}[htbp]
\caption{\label{tab:param}BABYSCAN parameters}
\small
\begin{tabular}{l|r}
\hline
Detector & $7.6 \times 12.7 \times 40.6$ cm  NaI,\\
& four identical units, \\
&each viewed by a 3" PMT\\
Shield thickness &bottom - 15\ cm\\
&top, side, rear - 10\ cm\\
&front - 13\ cm\\
Outer cover & GFRP\\
Inner cover & CFRP\\
Total weight & 5,700 kg\\
\hline
Measurement time & 4 min\\
Max.\ height of the subject & 130\ cm\\
Bed-detector distance& 20\ cm, 25\ cm, 30\ cm\\
%\hline
%$^{137}$Cs MDA & $< 50$  (Bq/body)\\
\hline
\end{tabular}
\end{table}

The detection efficiency can be further optimized by using a height-adjustable bed. The distance from the bed surface to the bottom of the NaI detector is either 20\ cm, 25\ cm or 30\ cm. The left panel of Fig.~\ref{fig:cutaway} shows a 20\ cm bed with a harness used for measuring small babies, while the right panel shows a child lying on a 25\ cm bed. The bed is pulled out when a child enters/leaves the measurement chamber, and it is pushed in during the measurement.

The size of the measurement chamber is 30\ cm (H) $\times 80$\ cm (W) $\times  140$\ cm (L), and the bed is 40\ cm (W) $\times 120$\ cm (L). These dimensions limit the maximum height of the subject to be about 130\ cm.

The measurement chamber and the detector compartment are surrounded by 10-cm-thick %SS400 
iron shielding as shown in the left panel of Fig.~\ref{fig:cutaway}. The shielding at bottom has an additional 5-cm of iron and the front has an additional 3-cm of iron. 
The size of the opening through which the subject enters the measurement chamber is 44\ cm (W) at the top and 32\ cm (W) at the bottom. This reflects an optimum balance between ease of use and maximum shielding of background radiation.

This iron structure is covered by an ergonomically designed plastic cover. 
The exterior surface of BABYSCAN is covered by smooth curved panels (made of glass-fiber reinforced plastic (GFRP)) colored with natural white for its gentle appearance.  In order to provide a cozy space for children, the interior surface is also covered by organic surface (made of carbon-fiber reinforced plastic (CFRP), so as to avoid the radium, thorium, and $^{40}$K background from the glass in GFRP), colored by light blue which looks like being made of soft materials. These panels are precisely assembled to eliminate the possibility of injuring baby's skin by their gaps or edges. 

All the materials used to manufacture BABYSCAN, including the bed and the tablet computer, were tested for natural radioactivity using a germanium detector prior to assembly.
Materials known previously to be problematic were avoided.  The approximate criteria for acceptance was when any radioactivity in the component would not cause a detectable peak in the background of the BABYSCAN detectors with a subject in place, when the subject is counted 10 times longer than normal, i.e. 30-40 minutes.  No artificial nuclides were detected, but occasionally, low levels of natural radioactivity were detected.

Table ~\ref{tab:param} summarizes the BABYSCAN parameters. 

\section{BABYSCAN calibration}
The BABYSCAN was calibrated using a Monte Carlo N-Particle Transport Code (MCNPX version 2.7.0)~\cite{mcnp} for a wide variety of weight and height combinations for each of the 3 bed-height positions.  These calibrations were validated with 1) a 4-year-old ANSI BOMAB phantom containing 290 kBq of $^{152}$Eu (made by  Japan Isotope Association), and 2)  2-year-old and 6-year-old ``universal'' phantoms, respectively containing 3113 Bq and 6225 Bq of $^{137}$Cs (made by STC RADEK, St.\ Petersburg), for three different bed heights.

In the ``universal'' phantom, polyethylene blocks and $^{137}$Cs-containing rods are combined to make six different  age and anthropometric types. We used 2- and 6-year-old phantoms to further check the BABYSCAN calibration, and also to compare the BABYSCAN's characteristics with those of FASTSCAN.

Table~\ref{tab:bomab} shows the results of these validation measurements at the $^{137}$Cs energy range.  The Phantom was counted and analyzed in the same manner that a child with that height would be analyzed.  The results are consistent with the claimed accuracy of the FASTSCAN ($\pm 20$\%).

\begin{table}
\caption{\label{tab:bomab} The results of validation measurements at the $^{137}$Cs energy range for the three phantoms, using different bed heights.}
\small
\begin{tabular}{l|c|ccc}
\hline
Phantom & Height (cm) & \multicolumn{3}{c}{Percent Difference}\\
& & 30\ cm & 25\ cm & 20\ cm\\
\hline
Block 2y & 83 & -1 & 3 & 4 \\
BOMAB 4y& 105 & -10 & -12 & -16  \\
Block 6y & 121 & -2 & -1 & -3 \\
\hline
\multicolumn{2}{r|}{Average} & -5.5 & -4.5 & -6 \\
\hline
\end{tabular}
\end{table}

\begin{figure}
\includegraphics[width=\columnwidth]{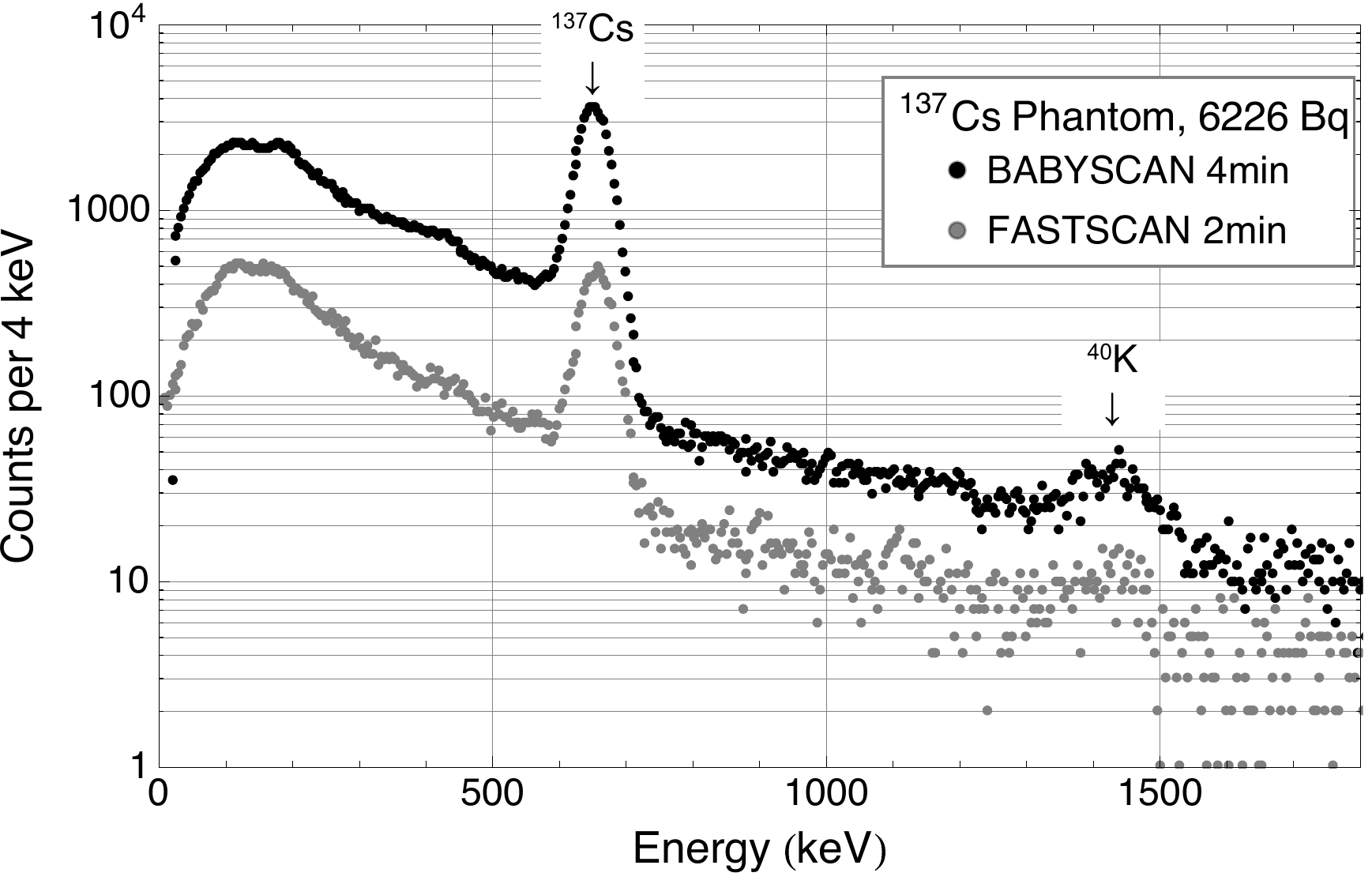}
\caption{\label{fig:compare} Comparison of the $^{137}$Cs 6-year-old phantom spectra. Black: BABYSCAN 4-minutes, Gray: FASTSCAN 2-minutes.}
\end{figure}

Fig.~\ref{fig:compare} shows the spectra of the 6-year-old phantom (6226 Bq of $^{137}$Cs) measured with the BABYSCAN (4-minutes, shown in black) and the FASTSCAN (2-minutes, shown in gray); they are installed in the same room of the hospital.  

The  $^{137}$Cs peak count of BABYSCAN is about 8 times larger as compared to the FASTSCAN. From the increase in the number of detectors ($\times 2$)  and in the measurement time ($\times 2$), one would na\"ively expect an increase of factor 4; the extra factor of 2 comes from the closer subject-detector distance than in the FASTSCAN, and because the detectors are closer to each other than in the FASTSCAN detector geometry.

The Cs-region background count of BABYSCAN is about 3.5 times higher than that of FASTSCAN, which is 13\% smaller than the factor 4 expected from the differences in both the number of detectors and the measurement time. This 13\% background reduction (despite a rather large opening at the top) is the result of the increased shielding.

\section{Initial operating experience of the BABYSCAN}

\begin{figure}
\includegraphics[width=0.8\columnwidth]{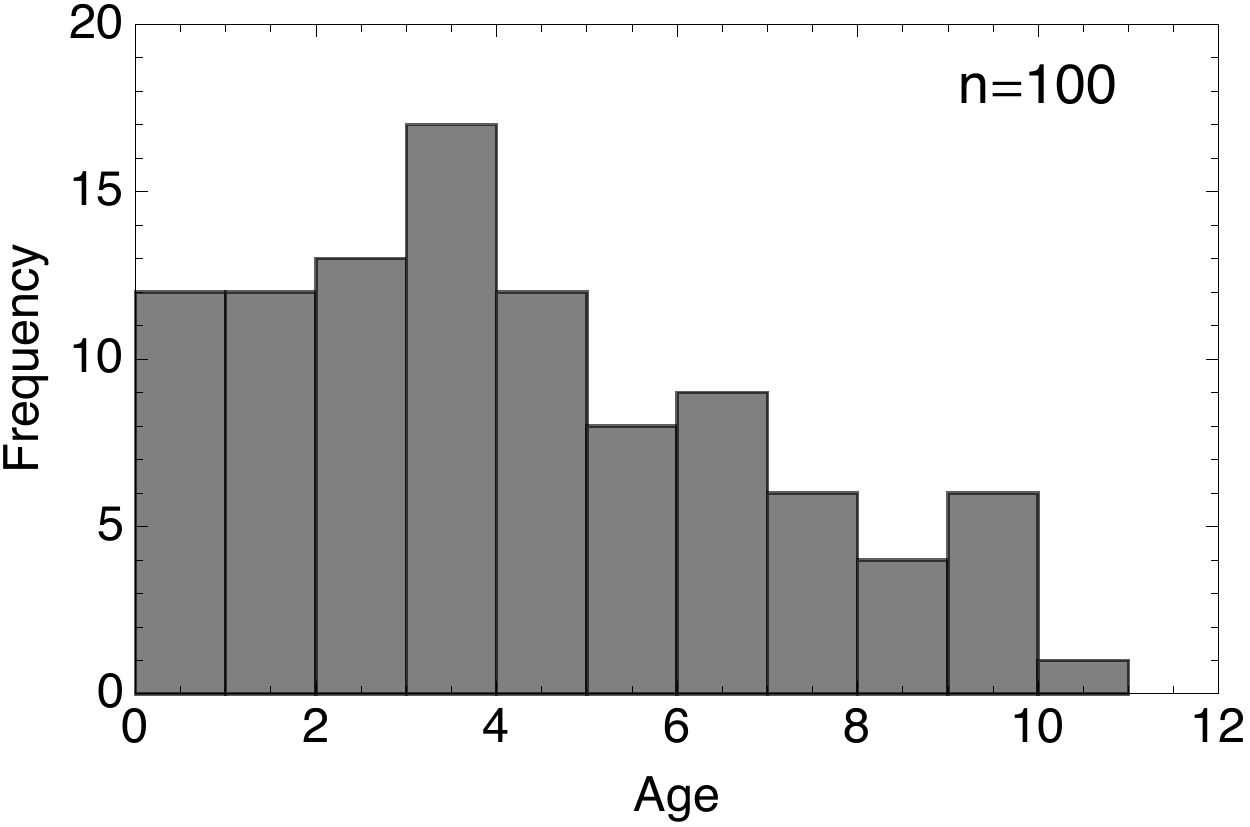}
\caption{\label{fig:age} Age distribution of the subjects.}
\end{figure}

\begin{figure}
\includegraphics[width=0.8\columnwidth]{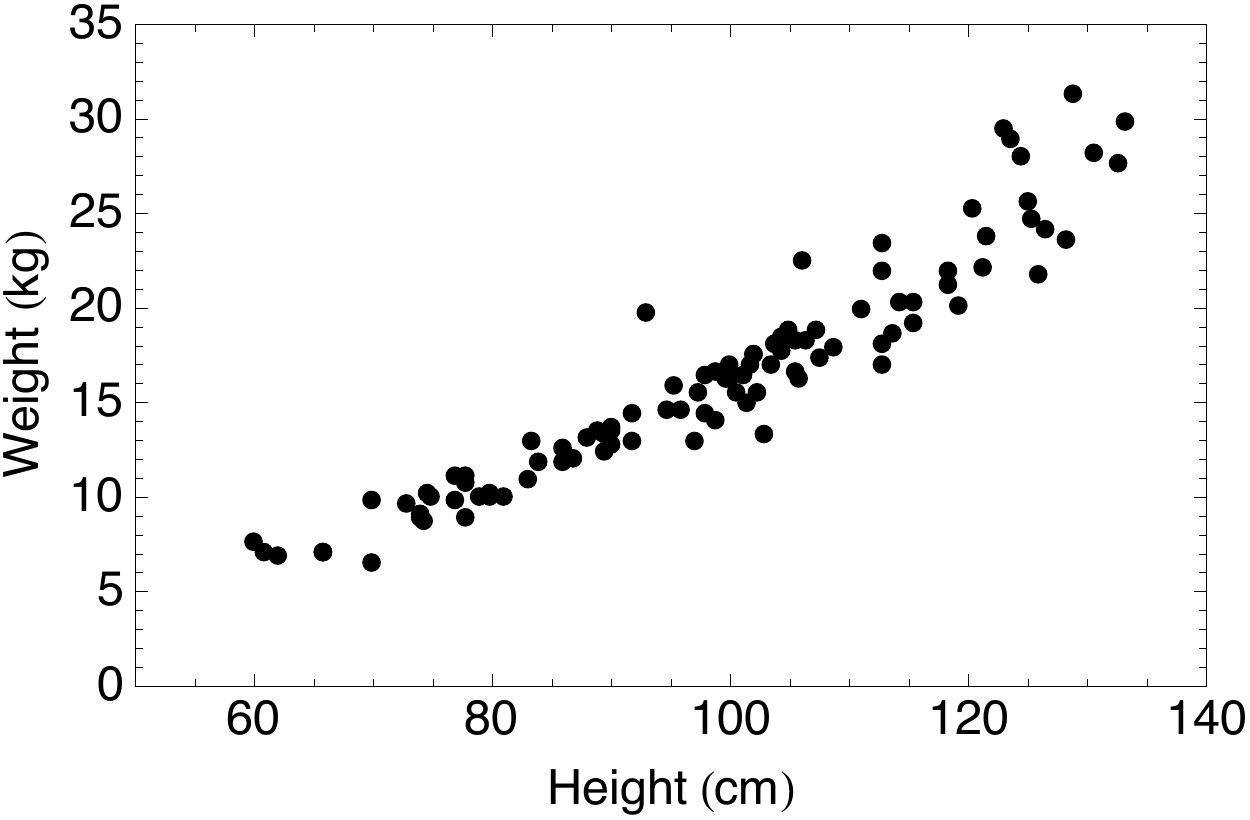}
\caption{\label{fig:height} Height vs weight of the subjects.}
\end{figure}

The first BABYSCAN unit, installed at the Hirata Central Hospital in Fukushima Prefecture, started operation on December 2, 2013. We here demonstrate its performance based on the data of first 100 subjects, whose age distribution (minimum 3.8 months old, maximum 10 year old, mean 4.2 year old) is shown in Fig.~\ref{fig:age}, and their anthropometric parameters are plotted in Fig.~\ref{fig:height} (minimum weight 6.5 kg, maximum weight 31.3\ kg, mean 16.1\ kg, minimum height 60.0\ cm, maximum height 133.3\ cm, mean 98.2\ cm).  This study was approved by the Ethics Committee of the University of Tokyo. 

\begin{figure}
\includegraphics[width=\columnwidth]{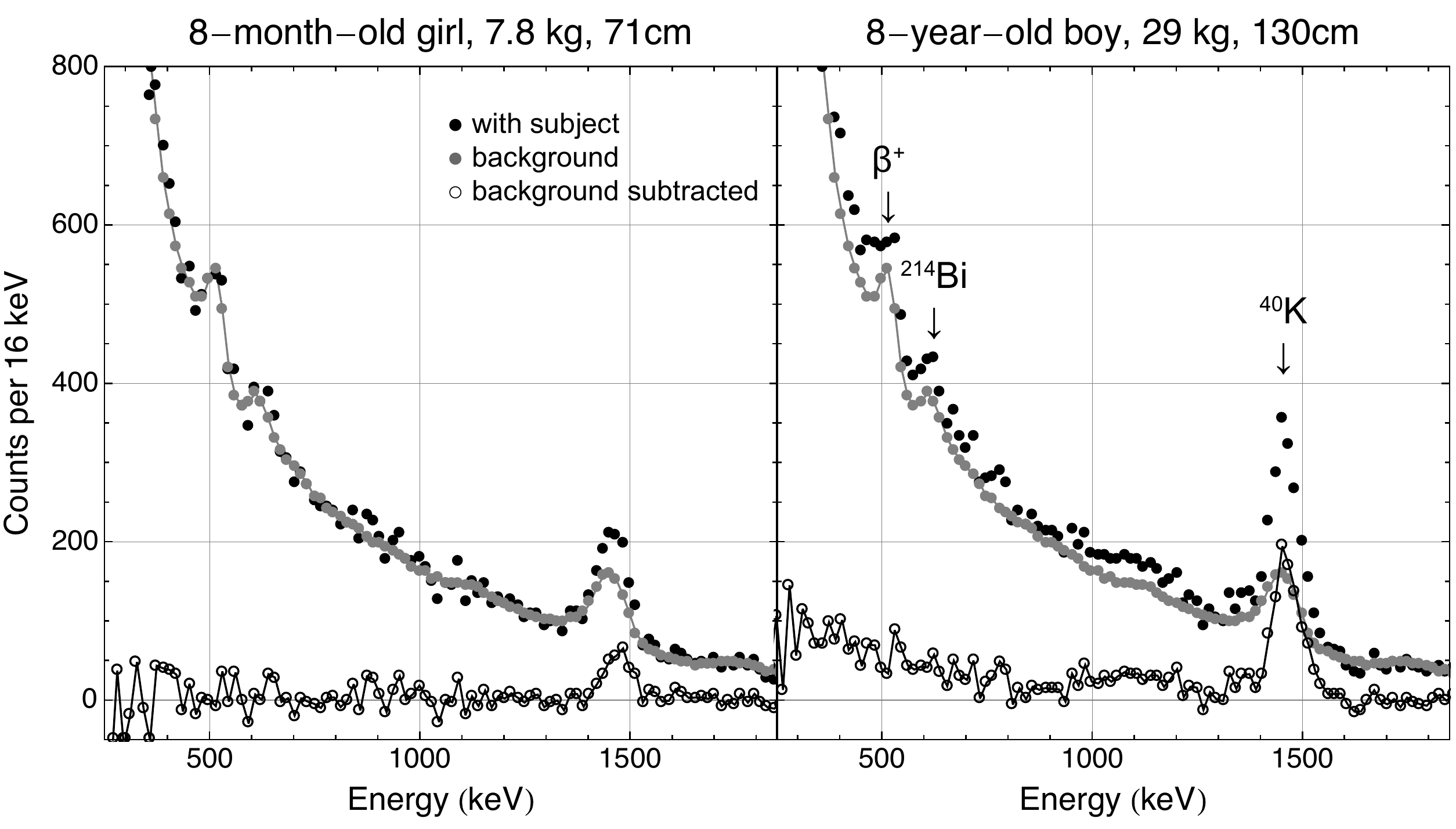}
\caption{\label{fig:spectra} Typical gamma-ray energy spectra measured with the BABYSCAN. Left: 8-month-old girl, right: 8-year-old boy. The spectra shown in black dots were taken with subjects (4 minutes), and those in gray dots were taken without subject (measured for 5 hours, normalized to 4 minutes). The background-subtracted spectra are shown in open circles.}
\end{figure}

\begin{figure}
\includegraphics[width=\columnwidth]{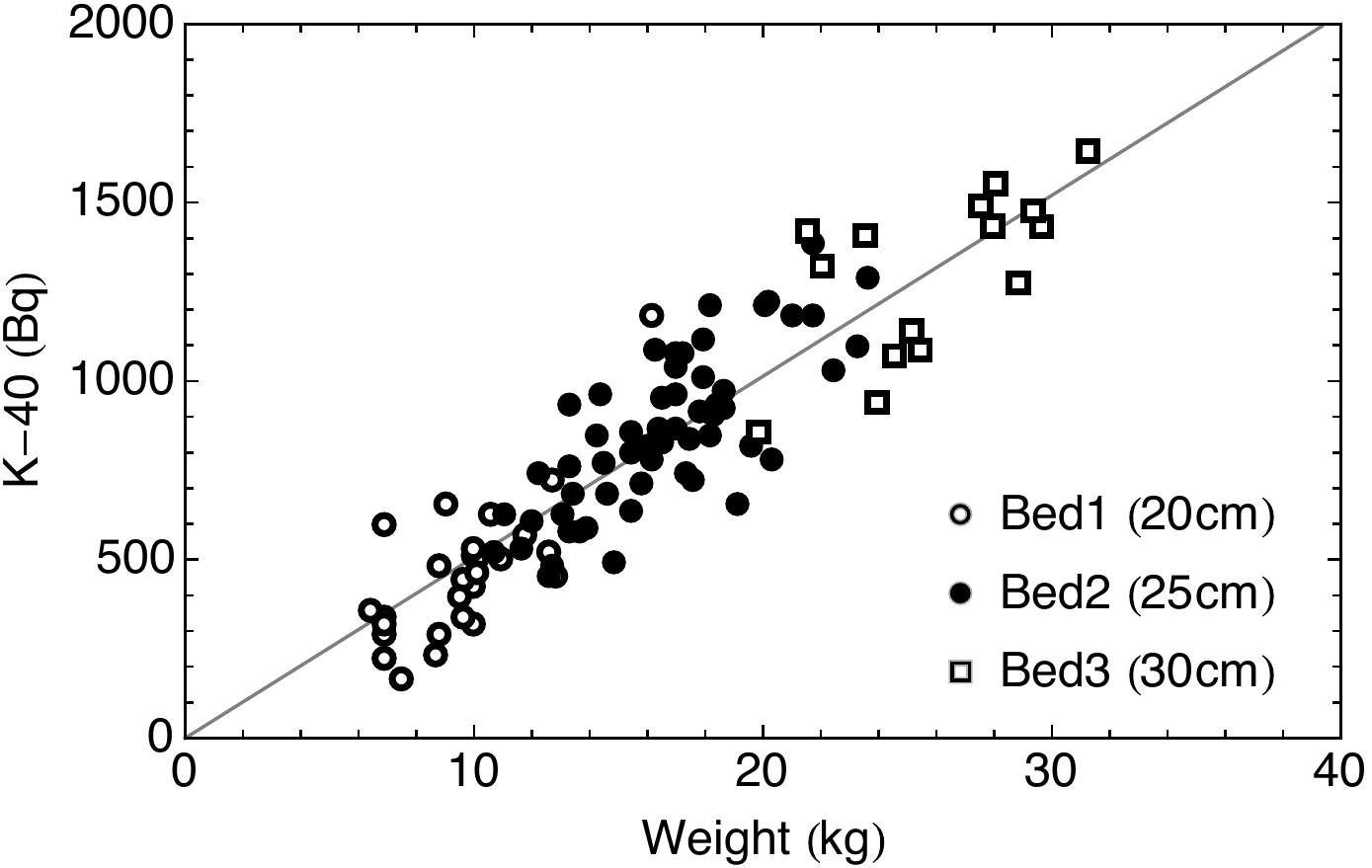}
\caption{\label{fig:k40} The distribution of the $^{40}$K activity (Bq/body) measured with the BABYSCAN vs the weight of the subject.}
\end{figure}

Radiocesium was not detected in any of the 100 subjects, and the mothers were happy to learn the test results.
Nevertheless, as expected, $^{40}$K was detected in all subjects. Typical gamma-ray energy spectra are presented in Fig.~\ref{fig:spectra}; the spectra shown in black dots were taken with subjects (4 minutes), and those shown in gray dots were taken without subject (measured for 5 hours, normalized to 4 minutes). A statistically-significant $^{40}$K peak was found in each of the background-subtracted spectrum (open circles).

Fig.~\ref{fig:k40} shows the distribution of the $^{40}$K activity (Bq/body) vs the weight of the subject. The data shown in open circles/filled circles/open squares were measured with the 20-cm/25-cm/30-cm bed. The data points show a linear correlation between the weight and the amount of $^{40}$K in the body, with a slope  of $50.7 \pm 0.9$ Bq kg$^{-1}$. This is consistent with the known amount of $^{40}$K in human body.

The data shown in Fig.~\ref{fig:k40} indicates that $^{40}$K activities as low as 300 Bq were reliably detected, and accurately measured.  One can use this information, to estimate the amount of $^{137}$Cs that can be reliably measured.  The gamma yield of $^{137}$Cs is 85\% while the gamma yield of $^{40}$K is only 10\%.  The $^{137}$Cs background [counts per keV] is 3 times the background of $^{40}$K, but the detector full width at half maximum (FWHM) for $^{137}$Cs is 0.6 times that of the $^{40}$K.  The efficiency at $^{137}$Cs energies is 1.4 times larger than at $^{40}$K.  These combine to convert a reliably measured 300 Bq $^{40}$K value into an estimated 35 Bq $^{137}$Cs value that should be reliably measured, all other things being equal.  This is consistent with the calculated $^{137}$Cs MDA described in the next paragraph.

\begin{figure}
\includegraphics[width=\columnwidth]{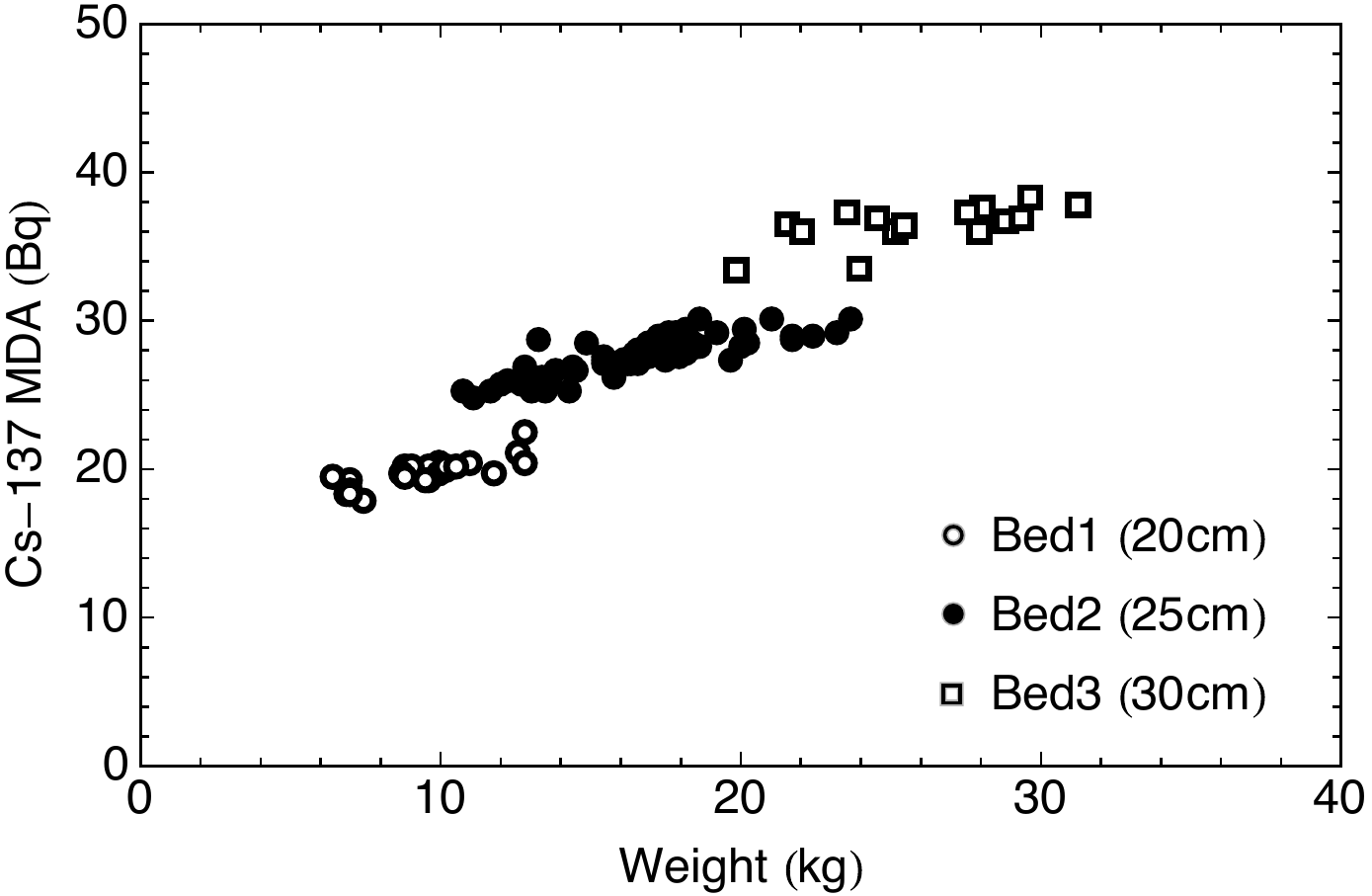}
\caption{\label{fig:mda}The minimum detectable activity (MDA) for $^{137}$Cs (Bq/body) vs subjects' weight.}
\end{figure}

The minimum detectable activity (MDA)~\cite{currie} for $^{137}$Cs (Bq/body), calculated for each subject, is plotted in Fig.~\ref{fig:mda} against weight (kg). Here again, data taken with 20-cm/25-cm/30-cm beds are shown in open circles/filled circles/open squares. As the bed-to-detector distance decreases, the solid angle increases and hence the MDA decreases. This plot clearly shows that our initial goal of achieving a detection limit lower than 50 Bq/body has been met.

\section{Conclusion}
BABYSCAN, a whole body counter for small children was developed, and the first unit has been installed at a hospital in Fukushima. The radiocesium detection limit of BABYSCAN is better than 50 Bq/body, which has been realized by a careful ergonomic design, optimized detector geometry and reinforced shielding. Even with this low detection limit,  radiocesium was not detected in any of the first 100 Fukushima children, while, as expected, $^{40}$K was detected in all subjects. The results of larger-scale measurements with the BABYSCAN will be reported in our forthcoming publications.

\begin{acknowledgments}
The authors would like to express their appreciation to Kinya Tagawa and Hisato Ogata of Takram design engineering for their assistance in the design of BABYSCAN. Thanks are also due to  
Nichinan Corporation and Shinwa kougyo Co.,Ltd.\ respectively for their contribution to the external panel and iron structure design and fabrication.
This work was partially supported by donations by many individuals to RH through The University of Tokyo Foundation.
\end{acknowledgments}

\end{document}